\documentclass[aps,superscriptaddress, twocolumn]{revtex4-1}
\usepackage[english]{babel}
\usepackage[ansinew]{inputenc}
\usepackage{times}
\usepackage{graphicx}
\usepackage{graphics}
\usepackage{amsmath}
\usepackage{amsfonts}
\usepackage{amssymb}
\usepackage{dsfont}
\usepackage{epstopdf}
\usepackage{makeidx}
\usepackage{subfigure}
\usepackage{color}
\usepackage{pgf}
\usepackage{bm}
\usepackage{tikz} \usepackage[normalem]{ulem}

\newcommand{\be}{\begin{equation}}
\newcommand{\ee}{\end{equation}}
\newcommand{\bea}{\begin{eqnarray}}
\newcommand{\eea}{\end{eqnarray}}

\newcommand\redsout{\bgroup\markoverwith{\textcolor{red}{\rule[0.5ex]{4pt}{0.8pt}}}\ULon}

\makeindex
\begin{document}


\title{Nonradiative emission and absorption rates of quantum emitters embedded in metallic systems: microscopic description and their determination from electronic transport}

\author{M. B. Silva Neto} 

\affiliation{Instituto de Fisica, Universidade Federal do Rio de Janeiro, Caixa Postal
68528, Rio de Janeiro, Brazil}

\author{F. M. D'Angelis} 

\affiliation{Instituto de Fisica, Universidade Federal do Rio de Janeiro, Caixa Postal
68528, Rio de Janeiro, Brazil}

\author{P. P. P. Foster} 

\affiliation{Instituto de Fisica, Universidade Federal do Rio de Janeiro, Caixa Postal
68528, Rio de Janeiro, Brazil}

\author{F. A. Pinheiro} 

\affiliation{Instituto de Fisica, Universidade Federal do Rio de Janeiro, Caixa Postal
68528, Rio de Janeiro, Brazil}

\begin{abstract}
We investigate nonradiative emission and absorption rates of two-level quantum
emitters embedded in a metal at low temperatures. We obtain the expressions for both nonradiative transition rates
and identify a unique, experimentally accessible way to obtain the nonradiative decay rates via electronic transport in the host metallic system. Our findings not 
only provide a microscopic description of nonradiative decay channels in metals, but they also allows one to identify and 
differentiate them from other decay channels, which is crucial to understand and control light-matter interactions at the nanoscale.

\end{abstract}

\maketitle

\section{Introduction}

Controlling and understanding light-matter interactions at the nanoscale is key for a broad range
of applications, including biosensing, imaging, and quantum information processing. Among the several processes that
govern light-matter interactions, spontaneous emission from Quantum Emitters (QE) (atoms, molecules, and quantum dots) is one of the most important for 
applications in nanophotonics. This radiative process strongly depends on the electromagnetic environment of the QE, as discovered in the pioneering work by Purcell~\cite{Purcell-46}, and it has 
been extensively investigated in several photonic systems, such as photonic cavities~\cite{bjork1991,gerard1998}, planar interfaces~\cite{urbach1998,johansen2008}, photonic crystals~\cite{yablonovitch1987,lodahl2004}, metamaterials~\cite{klimov2002,hyperbolicreview,jacob2012,kortkamp2013}, and waveguides~\cite{chen2010,kumar2013}. 

\begin{figure}
\begin{centering}
\includegraphics[width = 3.in]{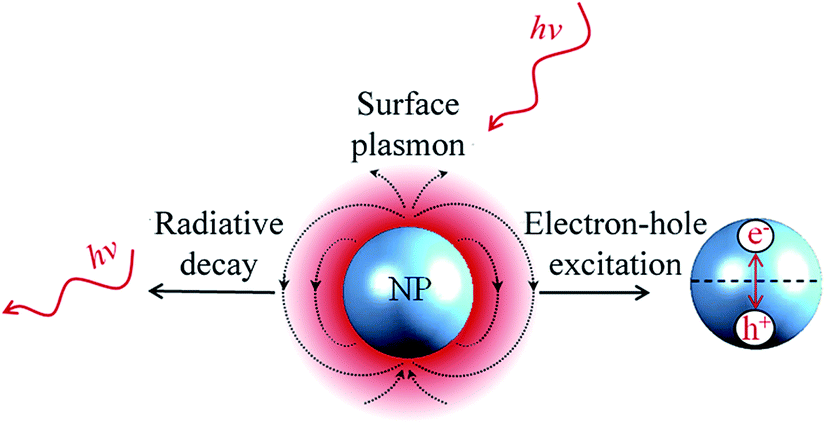}
\par\end{centering}
\label{Non-Radiative-Decay-Figure}
\caption{Possible relaxation mechanisms for a nanoparticle
(NP) bearing an electric dipole moment (thin field lines around
the NP) embedded in a metal. Besides photoemission (left) and 
surface plasmon absorption (top) transitions among the low (blue) 
and high (red) energy states of the electric-dipole moment in the 
nanoparticle can be induced by the inelastic scattering of electrons 
close to the Fermi surface (right).}
\end{figure}

In addition to the radiative relaxation, when the QE is placed near or inside metallic structures other decay pathways are available,
see Fig.~(\ref{Non-Radiative-Decay-Figure}). In this case the energy of the QE can be dissipated in a plasmonic channel. 
For instance, the proximity of a QE to metal-dielectric interfaces facilitates the excitation of surface plasmon polaritons, 
electromagnetic excitations related to the charge density waves on the surface of the metallic structure. This mechanism 
leads to a strong confinement of the electromagnetic field at metal-dielectric interfaces, which is the basis of many 
applications to enhance light-matter interactions, such as single optical plasmon generation~\cite{akimov2007,chang2006}, 
single molecule detection with surface-enhanced Raman scattering~\cite{nie1997}, and nanoantenna modified spontaneous emission~\cite{taminiau2008}.

Nonradiative relaxation is another decay pathway, where the QE energy can be dissipated via coupling to phonons, 
resistive heating, or quenching by other quantum emitters. Nonradiative relaxation is particularly important in metallic 
systems, where emission quenching may occur due to unavoidable dissipation even in systems with high spontaneous 
emission rate. In many cases of practical interest increasing the ratio between radiative and nonradiative decay channels 
is of great importance since the former actually determines the efficiencies of photonic devices, such as LEDs~\cite{fan1997}, 
and single-photon sources~\cite{eisaman2011}. In other situations it is very important to identify the nonradiative 
mechanism, distinguishing it from the plasmonic channel as it is the case of applications involving the excitation of 
single plasmon polaritons and subsequent controlled coupling between metallic nanowires~\cite{kumar2014,pors2015}.

In the present paper we identify a unique, experimentally accessible way to identity the nonradiative contribution 
to the total decay of QE inside metals. The situation in which the QE is embedded in metallic systems is within the 
reach of current nanofabrication techniques and is of increasing importance in applications involving 
metamaterials~\cite{roth2017}. By means of a microscopic, analytical approach, we compute the nonradiative decay 
channel of a QE embedded in a metal, in which dissipation is due to inelastic scattering of electrons close to the Fermi surface,
see Fig.~(\ref{Non-Radiative-Decay-Figure}) \cite{figure1}. After computing the transition rates for both nonradiative emission and 
absorption, we demonstrate that such quantities can be directly determined by the knowledge of experimentally 
accessible transport quantities, such as the optical and ac-conductivity, and even the dc-resistivity. This result not 
only provides a microscopic description of nonradiative decay channels in metals, but also allows one to identify and 
differentiate it to other decay channels, which is crucial for the development of disruptive optoelectronic plasmonic applications. 

%
%

This paper is organized as follows. In Sec. II we describe the methodology to microscopically calculate 
the transition rates for both nonradiative emission and absorption. In Sec. III we discuss and analyze the 
behavior of the decay rates as a function of the temperature, whereas Sec. IV is devoted to the conclusions.

\section{Methodology}

We consider a two-level QE embedded in a metal in which relaxation can occur 
via the inelastic scattering of electrons close to the Fermi surface, as it is schematically 
illustrated at the right side of Fig.~(\ref{Non-Radiative-Decay-Figure}). 
To model the coupling between the QD with the band electrons in a metal, let us considering 
the following interaction Hamiltonian: 



%
\be
H=\hbar\omega_{0}|e\rangle\langle e|+\sum_{k}\epsilon_{k}c_{k}^{\dagger}c_{k}+H_{int},
\ee
where $\hbar\omega_0$ is the energy splitting of the two level system, $\epsilon_{k}$
is the band dispersion relation for the electrons in the metal, $c_{k}^{\dagger}$ and
$c_{k}$ are creation and anihilation operators such that
\begin{eqnarray}
c_{k}\left|FS\right\rangle  & = & \sqrt{f^0\left(\epsilon_{k}\right)}\left|h_{k}\right\rangle,\nonumber\\
c_{k}^{\dagger}\left|FS\right\rangle  & = & \sqrt{1-f^0\left(\epsilon_{k}\right)}\left|p_{k}\right\rangle,
\label{fermionic-algebra}
\end{eqnarray}
with $\left|FS\right\rangle $ representing the Fermi sea, $\left|h_{k}\right\rangle $
representing the hole state, $\left|p_{k}\right\rangle $ representing
the particle state, and with $f^{0}\left(\epsilon_{k}\right)$ being the
Fermi-Dirac occupation probability. For simplicity, we shall omit spin indices
since we will be considering solely spin preserving scattering processes.
The interaction part of the Hamiltonian reads
\be
H_{int}=\sum_{\ell,m=g,e}\sum_{k,k^{\prime}}c_{k^{\prime}}^{\dagger}c_{k}
S\left(k^{\prime}-k\right)\tilde{V}_{QD}^{\ell m}(k^{\prime}-k)|\ell\rangle\langle m|,
\label{interaction-hamiltonian}
\ee
and describes the electrostatic interaction between an electronic charge density
for electrons in a metal and the potential generated by the two level system. Here 
$|g\rangle\langle g|$ and $|e\rangle\langle e|$ are the projection operators onto the 
ground, $|g\rangle$, and excited, $|e\rangle$, states, while $|e\rangle\langle g|$ 
and $|g\rangle\langle e|$ describe the electronic tunneling between the two levels. 
We have also introduced the impurity structure factor
\begin{equation}
S\left(k^{\prime}-k\right)=\sum_{{\bf R}_i}P_{R_{i}}e^{i(k^{\prime}-k)\cdot R_{i}},
\end{equation}
where $P_{R_{i}}$ gives the probability of having an emitter at ${\bf R}_i$.

In what follows we shall be interested in calculating the nonradiative emission
and absorption rates and, for this reason, we will restrict our calculations to the
$e\rightarrow g$ and $g\rightarrow e$ processes only, both associated to the 
matrix element (within the dipole approximation of the electrostatic Coulomb potential)
\begin{equation}
\widetilde{V}_{QD}^{ge}(k^{\prime}-k)=-\frac{i}{V}\frac{e}{\varepsilon_0\varepsilon_{r}(k-k^{\prime})}
\frac{1}{\left|k-k^{\prime}\right|}\hat{\xi}_{k-k^{\prime}}\cdot\left\langle g\left|\vec{\mu}\right|e\right\rangle ,
\end{equation}
describing the electron-atom coupling, where $e$ is the electric charge, $V$ is
the volume, $\varepsilon_0$ is the vacuum dielectric constant, $\varepsilon_{r}(k-k^{\prime})$ 
is the relative permissivity in the medium, $\vec{\mu}$ is the electric dipole moment of the emitter and
\begin{equation}
\hat{\xi}_{k-k^{\prime}}=\frac{k-k^{\prime}}{\left|k-k^{\prime}\right|}
\end{equation}
is the unit vector along the direction of $k-k^{\prime}$. For future
purposes it will be interesting to observe that the interacting part
of the Hamiltonian has the general structure
\begin{equation}
H_{int}=\sum_{k,k^{\prime}}c_{k^{\prime}}^{\dagger}c_{k}\tilde{V}_{QD}^{ge}(k^{\prime}-k)S\left(k^{\prime}-k\right)|g\rangle\langle e|,
\end{equation}
and that the eigenstates of the free Hamiltonian are written as
\begin{equation}
\left| \ell,FS \right> =\left| \ell \right> \otimes \left| FS \right>,
\end{equation}
where $\left|\ell\right> $ with $\ell=e,g$ are the eigenvectors
of the TLS Hamiltonian, and $\left|FS\right> $ are the eigenvectors
of the electron Hamiltonian in the number operator representation.

\section{Results and discussions}
\subsection{General structure for the non radiative decay rate}

%
%
For the above interaction Hamiltonian, with initial and final states
corresponding to
\begin{equation}
I\equiv\left|e,FS\right\rangle =\left|e\right\rangle \otimes\left|FS\right\rangle \rightarrow\left|g,h_{k}p_{k^{\prime}}\right\rangle =\left|g\right\rangle \otimes\left|h_{k}p_{k^{\prime}}\right\rangle \equiv F,
\end{equation}
where, $\ell,m=g,e$, and with energies
\begin{eqnarray*}
E_{I} & = & \hbar\omega_{0},\\
E_{F} & = & \epsilon_{k^{\prime}}-\epsilon_{k},
\end{eqnarray*}
the transition amplitude can be calculated from Fermi's golden rule 
\begin{eqnarray*}
\Gamma\left(k,k^{\prime};e,g\right) & = & \left(\frac{2\pi}{\hbar}\right)\left|\left\langle F\right|c_{k^{\prime}}^{\dagger}c_{k}\widetilde{V}_{QD}^{ge}(k^{\prime}-k)|g\rangle\langle e|\left|I\right\rangle \right|^{2}\\
 & \times & \left\langle S\left(k^{\prime}-k\right)S\left(k^{\prime}-k\right)\right\rangle \delta\left(E_{F}-E_{I}\right).
\end{eqnarray*}
The matrix element can be calculated with the use of the fermionic
algebra (\ref{fermionic-algebra}) and from the fact that
$\left\langle S\left(k^{\prime}-k\right)S\left(k^{\prime}-k\right)\right\rangle=N_{imp}$,
for a small number of dilute impurities (emitters). We are now ready to rewrite 
the transition amplitude in terms of the initial and final momentum states, $\left|h_{k}\right\rangle$
representing the hole state, $\left|p_{k^{\prime}}\right\rangle$
representing the electron state, and in terms of the occupation probabilities 
for the ground, $n_g$, and excited, $n_e$, states in the two-level system 
\bea
\Gamma\left(k,k^{\prime};e,g\right)&=&(\frac{2\pi}{\hbar})N_{imp} f^{0}\left(\epsilon_{k}\right)
[1-f^{0}\left(\epsilon_{k^\prime}\right)]n_{e}(1-n_{g})\nonumber\\
&\times&|\widetilde{V}_{QD}^{ge}(k^{\prime}-k)|^{2}\delta(\epsilon_{k^{\prime}}-\epsilon_{k}-\hbar\omega_{0}).
\eea

The next step is to calculate the relaxation rates through summing
up transition amplitudes using the relation
\begin{equation}
\sum_{k,\sigma}\rightarrow V\times2\times\int\frac{d\epsilon_{k}}{\hbar v_{k}}\int\frac{d\Omega_{k}}{\left(2\pi\right)^{3}},
\end{equation}
where the factor $2$ accounts for spin degeneracy, $v_{k}=\hbar k/m^*$ is
the velocity for a nearly free electron, parabolic band approximation with
effective mass $m^*$, and
\begin{equation}
d\Omega_{k}=k^{2}d\Omega=k^{2}\sin\varphi d\theta d\varphi.
\end{equation}
For the nonradiative (nr) decay from the excited to ground states
we thus have
\begin{eqnarray*}
\Gamma^{e\rightarrow g}_{nr}(T) & = & \sum_{k,k^{\prime},\sigma,\sigma^{\prime}}\Gamma\left(k,k^{\prime};e,g\right)\\
 & = & V^{2}\times 4 \times\left(\frac{2\pi}{\hbar}\right)N_{imp}n_{e}(1-n_{g}) \nonumber\\
 & \times & \left\{ \int\frac{d\epsilon_{k}}{\hbar v_{k}}\int\frac{d\epsilon_{k^{\prime}}}{\hbar v_{k^{\prime}}}f^{0}(\epsilon_{k})[1-f^{0}(\epsilon_{k^{\prime}})]\right\} \\
 & \times & \left\{ \int\frac{d\Omega_{k}}{\left(2\pi\right)^{3}}\int\frac{d\Omega_{k^{\prime}}}{\left(2\pi\right)^{3}}\left|\widetilde{V}_{QD}^{ge}(k^{\prime}-k)\right|^{2}\right\} \\
 & = & \left(\frac{2\pi}{\hbar}\right)\left(\frac{m^{*}k_{F}}{\pi^{2}\hbar^{2}}\right)^{2}N_{imp}n_{e}(1-n_{g})\left[\frac{\hbar\omega_{0}}{1-e^{-\beta\hbar\omega_{0}}}\right]\\
 & \times & \frac{e^{2}}{\varepsilon_{0}^{2}}\frac{\mu^{2}}{3}\left[\int\frac{d\Omega}{4\pi}\int\frac{d\Omega^{\prime}}{4\pi}\frac{\left|k-k^{\prime}\right|^{2}}{(\left|k-k^{\prime}\right|^{2}+\lambda_{TF}^{-2})^2}\right].
\end{eqnarray*}
We identify the electronic density of states at the Fermi level
\be
N(\epsilon_{F})=\frac{m^{*}k_{F}}{\pi^{2}\hbar^{2}}
\ee
and we have defined the quantity
\begin{equation}
g_{nr}^{2}=\frac{e^{2}}{\varepsilon_{0}^{2}}\frac{\mu^{2}}{3}\frac{\hbar\omega_{0}}{k_F^2}
\left[\frac{1}{1+2k_F^2\lambda_{TF}^2}+\ln{\left(1+2k_F^2\lambda_{TF}^2\right)}-1\right],
\label{gnr0}
\end{equation}
in terms of the Thomas-Fermi screening length $\lambda_{TF}$. Now the
nonradiative decay rate is simply
\be
\Gamma^{e\rightarrow g}_{nr}(T)=\left(\frac{2\pi}{\hbar}\right)N^{2}\left(\epsilon_{F}\right)N_{imp}n_{e}(1-n_{g})\left[\frac{1}{1-e^{-\beta\hbar\omega_{0}}}\right]g_{nr}^{2}.
\label{gnrf}
\ee
To arrive at the above result we have used that, at low temperatures,
the function
\begin{equation}
f^{0}(\epsilon_{k})\left[1-f^{0}(\epsilon_{k}+\hbar\omega_{0})\right]
\end{equation}
is strongly peaked around the Fermi energy and thus we projected all
states $k\rightarrow k_{F}$ and $k^{\prime}\rightarrow k_{F}$. Furthermore,
we calculated
\begin{equation}
\int d\epsilon_{k}f^{0}(\epsilon_{k})[1-f^{0}(\epsilon_{k}+\hbar\omega_{0})]=\frac{\hbar\omega_{0}}{1-e^{-\beta\hbar\omega_{0}}}.
\end{equation}

\subsection{General structure for the non radiative absorption rate}

Similarly to the result obtained above, the nonradiative absorption
rate is given by
\begin{equation}
\Gamma^{g\rightarrow e}_{nr}(T)=
\left(\frac{2\pi}{\hbar}\right)N^{2}\left(\epsilon_{F}\right)N_{imp}n_{g}\left(1-n_{e}\right)\left[\frac{1}{e^{\beta\hbar\omega_{0}}-1}\right]g_{nr}^2,
\label{gnra}
\end{equation}
where $n_{e}$ and $n_{g}$ have been interchanged and the thermal
factor is also different
\begin{equation}
\left[\frac{1}{1-e^{-\beta\hbar\omega_{0}}}\right]_{decay}\rightarrow\left[\frac{1}{e^{\beta\hbar\omega_{0}}-1}\right]_{absorption},
\end{equation}
satisfying detailed balance.

\subsection{Normalized decay and absorption rates}

\begin{figure}
\begin{centering}
\includegraphics[width = 3.in]{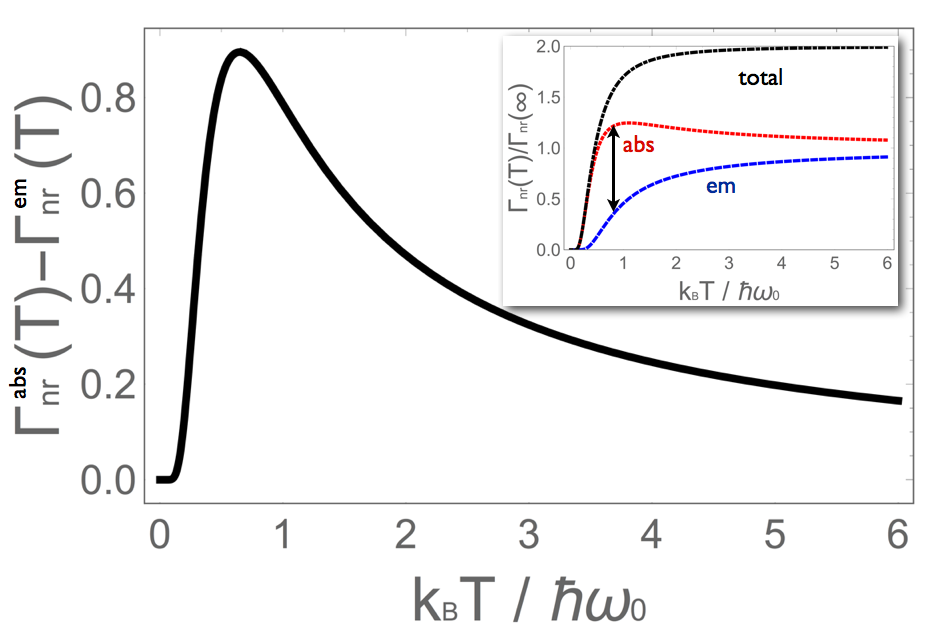}
\par\end{centering}
\caption{Main panel: black (solid) curve representing the relative 
difference between absorption and emission rates for non radiative 
transitions for an emitter embedded in a low temperature metallic 
host. Inset: red (dotted) curve representing the absorption rate, blue 
(dashed) curve representing the emission rate, and black (solid) 
curve representing the sum of absorption and emission rates. The 
curve shown in the main panel measures the distance between the 
red (dotted) and blue (dashed) curves in the inset.}
\label{NR-Decay-Absorption-Rates}
\end{figure}

A meaningful quantity to be defined is the nonradiative decay rate
normalized by its saturation value
\begin{equation}
\Gamma^{sat}_{nr}\left(\hbar\omega_{0}\ll k_{B}T\ll E_{F}\right)=
\left(\frac{2\pi}{\hbar}\right)N_{imp}\frac{1}{4}N^{2}\left(\epsilon_{F}\right)\left(\frac{g_{nr}^2}{\beta\hbar\omega_0}\right),
\end{equation}
since, for $\hbar\omega_{0}\ll k_{B}T\ll E_{F}$, $n_{g}\rightarrow1/2$ and 
$n_{e}\rightarrow1/2$. With this 
definition we arrive at
\begin{equation}
\frac{\Gamma^{e\rightarrow g}_{nr}\left(T\right)}{\Gamma_{nr}^{sat}}=4\,n_{e}\left(1-n_{g}\right)\left[\frac{\beta\hbar\omega_{0}}{1-e^{-\beta\hbar\omega_{0}}}\right],
\end{equation}
which is a dimensionless number between $0$ and $1$. The same is
valid for the absorption rate
\begin{equation}
\frac{\Gamma^{g\rightarrow e}_{nr}\left(T\right)}{\Gamma_{nr}^{sat}}=4\,n_{g}\left(1-n_{e}\right)\left[\frac{\beta\hbar\omega_{0}}{e^{\beta\hbar\omega_{0}}-1}\right].
\end{equation}
The results for both nonradiative emission and absorption rates, as well
as their sum, are plotted in the inset of Fig. \ref{NR-Decay-Absorption-Rates}.
We can also study the difference between the nonradiative absorption and emission rates
\begin{equation}
\Delta\Gamma_{nr}\equiv\frac{\Gamma_{nr}^{g\rightarrow e}\left(T\right)-\Gamma_{nr}^{e\rightarrow g}\left(T\right)}{\Gamma_{nr}^{sat}},
\end{equation}
which is a quantity between $0$ and $1$ and is plotted as the
main panel in Fig. \ref{NR-Decay-Absorption-Rates}. 

%
%

Figure \ref{NR-Decay-Absorption-Rates} reveals that the absorption processes, in 
which a recombination of a particle and a hole provides energy for the
$g\rightarrow e$ transition, dominates for $k_B T\ll\hbar\omega_0$. Absorption rapidly 
increases with temperature, but eventually saturates due to the decrease 
in the thermal occupation of the ground state for $k_B T\gg\hbar\omega_0$. 
On the other hand, emission processes in which a particle and a hole are 
created, receiving energy from the $e\rightarrow g$ transition, are less 
frequent at all temperatures because of the lower thermal occupation in 
the excited state of the emitter. Nevertheless, it also increases with the
temperature and also saturates for $k_B T\gg\hbar\omega_0$. Remarkably,
let us point out that such large difference between the absorption and emission 
rates indicates that the phenomenon of {\it generation} is occurring at the
emitter due to the inelastic scattering from Fermi surface states at low 
temperatures. This result suggests one can prepare the emitter in excited state by setting the temperature around $k_B T \leq \hbar\omega_0$,
which may find potential applications in quantum nanophotonics.  

\section{Connection to electronic transport}

Let us now see how we can extract information about the non radiative transition
rates from dc- and magneto-transport experiments. We shall first look into the 
dc-transport, with an externally applied electric field, $E\neq 0$, where the resistivity 
\begin{equation}
\rho=\frac{m^{*}}{ne^{2}}\frac{1}{\tau_{tr}},
\end{equation}
is given in terms of the inverse transport lifetime, $1/\tau_{tr}$, with $m^*$ being
the effective mass, $n$ the electronic density and $e$ the electric charge.
The quantity of interest will be the inverse transport lifetime, $1/\tau_{tr}$,
which can be calculated using a variational approach to the linearized Boltzmann's 
transport equations within the relaxation time approximation \cite{allen}. 
Next, we shall focus on the 
magneto-transport, with externally applied electric, $E\neq 0$, and magnetic, $B\neq 0$,
fields, where the Shubnikov-de-Haas oscillations \cite{sdho,deltaRxx}
\begin{equation}
\Delta R_{xx}(\omega_c)=4 R_0 e^{-\pi/\omega_c\tau_{q}}\cos{\left(\frac{2 \hbar \pi^2 n}{m^* \omega_c}-\pi\right)}\chi(T),
\end{equation}
are given in terms of the quantum lifetime, $\tau_{q}$, with $R_0$ being
the zero field resistance, $\omega_c=eB/m^*$ the ciclotron frequency,
and $\chi(T)=(2\pi^2 k_BT/\hbar\omega_c)\sinh{(2\pi^2 k_BT/\hbar\omega_c)}$,
a the thermal damping factor. The quantity of interest here is the inverse 
quantum lifetime, $1/\tau_{q}$, which can also be calculated using a
variational procedure \cite{allen}. For both cases (dc- and magneto-
transport) we shall explain how the non radiative transition rates
calculated earlier can be extracted from experiments.

\subsection{Connection to transport lifetime $\tau_{tr}$}

According to the interaction Hamiltonian (\ref{interaction-hamiltonian}) there are four channels 
for scattering between the emitter and Fermi surface electrons: two elastic channels, 
$g\rightarrow g$ and $e\rightarrow e$, and two inelastic channels,
$g\rightarrow e$ (absorption) and $e\rightarrow g$ (emission). The inverse transport lifetime,
$1/\tau_{tr}$, can thus be calculated using Mathiessen's rule
\begin{equation}
\frac{1}{\tau_{tr}}=\frac{1}{\tau_{gg}}+\frac{1}{\tau_{ge}}+\frac{1}{\tau_{eg}}+\frac{1}{\tau_{ee}},
\end{equation}
in which each individual contribution, $1/\tau_{\ell m}$, to the total inverse scattering time can 
be calculated from a variational principle to the linearized Boltzmann's equations within the 
relaxation time approximation \cite{allen}
\begin{equation}
\frac{1}{\tau_{\ell m}}=\frac{1}{2k_{B}T}\frac{\sum_{k,k^{\prime}}\left[\vec{u}\cdot\left(\vec{v}_{k}-\vec{v}_{k^{\prime}}\right)\right]^{2}
P_{k^{\prime},k}^{\ell m}}{\sum_{k}\left(\vec{u}\cdot\vec{v}_{k}\right)^{2}\left(-\frac{\partial f_{k}^{0}}{\partial\epsilon_{k}}\right)},
\label{variational-principle}
\end{equation}
where $\vec{u}$ corresponds to the direction of the applied
electric field, and $P_{k^{\prime},k}^{\ell m}$ are the scattering 
amplitudes from ${\bf k}$ to ${\bf k}^\prime$, between states
labelled by $\ell,m=g,e$. 

The denominator in (\ref{variational-principle}) can be written as
\begin{equation}
\sum_{k,\sigma}\left(\vec{u}\cdot\vec{v}_{k}\right)^{2}\left(-\frac{\partial f_{k}^{0}}{\partial\epsilon_{k}}\right) 
= \frac{1}{3\pi^{2}}\frac{v_{F}}{\hbar}=\frac{n}{m^*},
\end{equation}
where $n$ is the electronic density of the Fermi system, 
and the factor $1/3$ arises from the spherical symemtry of the
problem. As for the numerator in (\ref{variational-principle}) we 
shall write
\begin{equation}
\langle P^{\ell m}\rangle_{tr}=\frac{1}{2k_{B}T}\sum_{k,k^{\prime}}\left[\vec{u}\cdot\left(\vec{v}_{k}-\vec{v}_{k^{\prime}}\right)\right]^{2}
P_{k^{\prime},k}^{\ell m},
\end{equation}
where the scattering amplitudes are
\begin{equation}
P_{k^{\prime},k}^{gg}=(\frac{2\pi}{\hbar})N_{imp}|\widetilde{V}_{QD}^{gg}(k^{\prime}-k)|^{2}\delta(\epsilon_{k}-\epsilon_{k^{\prime}})f_{k}^{0}(1-f_{k^{\prime}}^{0})n_{g},
\end{equation}
\begin{equation}
P_{k^{\prime},k}^{eg}=(\frac{2\pi}{\hbar})N_{imp}|\widetilde{V}_{QD}^{eg}(k^{\prime}-k)|^{2}\delta(\epsilon_{k}-\epsilon_{k^{\prime}}-\hbar\omega_{0})f_{k}^{0}(1-f_{k^{\prime}}^{0})n_{g}(1-n_{e}),
\end{equation}
\begin{equation}
P_{k^{\prime},k}^{ge}=(\frac{2\pi}{\hbar})N_{imp}|\widetilde{V}_{QD}^{ge}(k^{\prime}-k)|^{2}\delta(\epsilon_{k^{\prime}}-\epsilon_{k}-\hbar\omega_{0})f_{k}^{0}(1-f_{k^{\prime}}^{0})n_{e}(1-n_{g}),
\end{equation}
\begin{equation}
P_{k^{\prime},k}^{ee}=(\frac{2\pi}{\hbar})N_{imp}|\widetilde{V}_{QD}^{ee}(k^{\prime}-k)|^{2}\delta(\epsilon_{k}-\epsilon_{k^{\prime}})f_{k}^{0}(1-f_{k^{\prime}}^{0})n_{e}.
\end{equation}
In this case, for the two elastic scattering processes we find
\begin{eqnarray}
\langle P^{gg}\rangle_{tr} & = & \frac{4 V^2}{2k_{B}T}\int\frac{d^{3}k^{\prime}}{(2\pi)^{3}}\int\frac{d^{3}k}{(2\pi)^{3}}[\vec{u}\cdot(\vec{v}_{k}-\vec{v}_{k^{'}})]^2P_{k,k^{\prime}}^{gg} \nonumber \\
 & = & \left(\frac{2\pi}{\hbar}\right)N^{2}\left(\epsilon_{F}\right)N_{imp}n_{g} g_{el}^2,
\label{pgg} 
\end{eqnarray}
and
\begin{eqnarray}
\langle P^{ee}\rangle_{tr} & = & \frac{4 V^2}{2k_{B}T}\int\frac{d^{3}k^{\prime}}{(2\pi)^{3}}\int\frac{d^{3}k}{(2\pi)^{3}}[\vec{u}\cdot(\vec{v}_{k}-\vec{v}_{k^{'}})]^2 P_{k,k^{\prime}}^{ee}  \nonumber \\
 & = & \left(\frac{2\pi}{\hbar}\right)N^{2}\left(\epsilon_{F}\right)N_{imp}n_{e} g_{el}^2,
 \label{pee} 
\end{eqnarray}
where
\be
g_{el}^{2}=\frac{1}{6}\frac{e^{2}}{\varepsilon_{0}^{2}}\left(\frac{\hbar}{m^*}\right)^2\frac{1}{k_F^2}
\left[\frac{1}{1+2k_F^2\lambda_{TF}^2}+\ln{\left(1+2k_F^2\lambda_{TF}^2\right)}-1\right].
\ee

\begin{figure}
\begin{centering}
\includegraphics[width = 3.in]{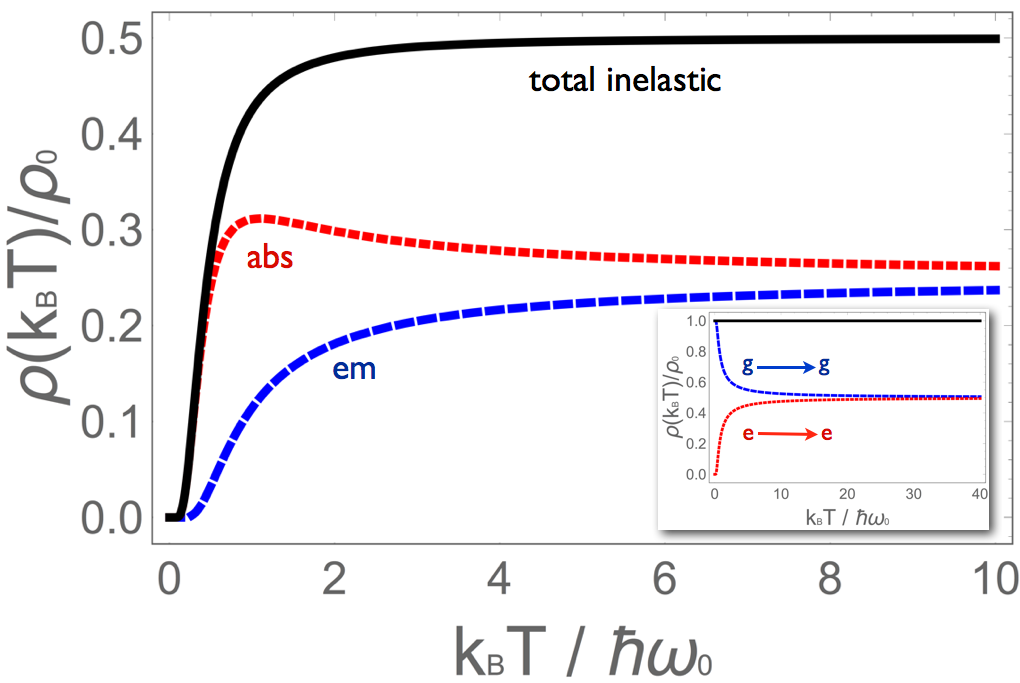}
\par\end{centering}
\caption{Temperature dependence of the dc-resistivity, $\rho(T)$,
normalized by its zero temperature value, $\rho_0$. Main pannel: inelastic channels,
red (dotted) representing absorption, $g\rightarrow e$, blue (dashed)
representing emission, $e\rightarrow g$, and black (solid) representing
the sum of the two inelastic (emission and absorption) contributions. 
Inset: elastic channels, red (dotted) representing $e\rightarrow e$ processes, 
blue (dashed) representing $g\rightarrow g$ processes, and black (solid) 
representing the sum of the two elastic contributions to transport.}
\label{Elastic-Scattering-Slow-Dot}
\end{figure}

The elastic contributions to the resistivity are shown in the inset of  
Fig. \ref{Elastic-Scattering-Slow-Dot}. As we can see, the sum of
the two elastic contributions is temperature independent mostly
because $n_g+n_e=1$ and $\tilde{V}^{gg}=\tilde{V}^{ee}$, even
though each of the two individual scattering channels exhibits a 
characteristic evolution with the temperature until saturation at
$n_e=n_g=1/2$ for $k_B T\gg\hbar\omega_0$. Furthermore, 
since this contribution has its origins in elastic processes, $1/\tau_{gg}$ 
and $1/\tau_{ee}$ do not contain information about the structure 
of the emitter, neither through the electric dipole moment, $\vec{\bf \mu}$,
nor through the characteristic energy of the emitter, $\hbar\omega_0$. 
Hence the elastic contributions to the resistivity have no connection to the transition rates calculated
earlier.

%
%

For the two inelastic scattering processes we have
\begin{eqnarray}
\langle P^{ge}\rangle_{tr} & = & \frac{4 V^2}{2k_{B}T}\int\frac{d^{3}k^{\prime}}{(2\pi)^{3}}\int\frac{d^{3}k}{(2\pi)^{3}}[\vec{u}\cdot(\vec{v}_{k}-\vec{v}_{k^{'}})]^2 P_{k,k^{\prime}}^{ge} \nonumber \\
  & = & \left(\frac{2\pi}{\hbar}\right)N^{2}\left(\epsilon_{F}\right)N_{imp}n_{g}(1-n_{e})\left[\frac{\beta}{e^{\beta\hbar\omega_{0}}-1}\right]g_{in}^2,
\label{pger}  
\end{eqnarray}
and
\begin{eqnarray}
\langle P^{eg}\rangle_{tr} & = & \frac{4 V^2}{2k_{B}T}\int\frac{d^{3}k^{\prime}}{(2\pi)^{3}}\int\frac{d^{3}k}{(2\pi)^{3}}[\vec{u}\cdot(\vec{v}_{k}-\vec{v}_{k^{'}})]^2 P_{k,k^{\prime}}^{eg} \nonumber \\
  & = & \left(\frac{2\pi}{\hbar}\right)N^{2}\left(\epsilon_{F}\right)N_{imp}n_{e}(1-n_{g})\left[\frac{\beta}{1-e^{-\beta\hbar\omega_{0}}}\right]g_{in}^2,
\label{pegr}  
\end{eqnarray}
where
\begin{eqnarray}
g_{in}^{2}&=&\frac{e^{2}}{\varepsilon_{0}^{2}}\frac{\mu^{2}}{3}\frac{\hbar\omega_{0}}{\lambda_{TF}^2}\left(\frac{\hbar}{m^*}\right)^2\nonumber\\
& \times & \left[\frac{2k_F^2\lambda_{TF}^2(1+k_F^2\lambda_{TF}^2)}{1+2k_F^2\lambda_{TF}^2}-\ln{\left(1+2k_F^2\lambda_{TF}^2\right)}\right].
\label{gin}
\end{eqnarray}
The inelastic contributions to the resistivity are shown in the main
panel of Fig. \ref{Elastic-Scattering-Slow-Dot}. In contrast to the elastic case, both the 
individual contributions to the resistivity, as well as their sum, 
have a strong temperature dependence that is analogous to
the temperature dependence of the normalized emission and 
absorption rates shown in Fig. \ref{NR-Decay-Absorption-Rates}.

Also differently to the elastic case, the inelastic contributions 
$1/\tau_{eg}$ and $1/\tau_{ge}$ do carry information about the structure of the emitter, through 
both the electric dipole moment, $\vec{\bf \mu}$, and the characteristic energy, $\hbar\omega_0$. 
Indeed, it is important to note the direct connection that exists between the non-radiative 
decay rate, given by (\ref{gnr0}) and (\ref{gnrf}), and the inelastic scattering process given 
by (\ref{pegr}). Both quantities are proportional to $\mu^2\hbar\omega_0$ and only differ 
by a constant factor that is fixed by the Fermi wave vector, $k_F$, and the Thomas-Fermi
screening length, $\lambda_{TF}$, see $g_{nr}^{2}$ in (\ref{gnr0}) and $g_{in}^{2}$ in (\ref{gin}). 
By the same token, a similar mapping exists between the nonradiative absorption rate (\ref{gnra}) 
and the other inelastic scattering process, given by (\ref{pger}). Again, these two quantities only 
differ by a constant factor.  Altogether these findings unveil the relation between nonradiative 
decay and absorption rates and the inelastic scattering processes, suggesting that one can 
obtain information about nonradiative decay by means of electronic transport.

\subsection{Connection to quantum lifetime $\tau_{q}$}

Similar to the case of the dc-resistivity, the quantum lifetime
can also be calculated, using Mathiessen's rule
\begin{equation}
\frac{1}{\tau_{q}}=\frac{1}{\tau_{gg}}+\frac{1}{\tau_{ge}}+\frac{1}{\tau_{eg}}+\frac{1}{\tau_{ee}},
\label{tauqp}
\end{equation}
where now each contribution corresponds to \cite{allen}
\begin{equation}
\frac{1}{\tau_{\ell m}}=\frac{1}{k_{B}T}\frac{\sum_{k,k^{\prime}}
P_{k^{\prime},k}^{\ell m}}{\sum_{k}\left(-\frac{\partial f_{k}^{0}}{\partial\epsilon_{k}}\right)}.
\label{quantum-lifetime}
\end{equation}
Again, for the numerator in (\ref{quantum-lifetime}) we define
\begin{equation}
\langle P^{\ell m}\rangle_{q}=\frac{1}{k_{B}T}\sum_{k,k^{\prime}}P_{k^{\prime},k}^{\ell m}.
\end{equation}
For the two inelastic scattering processes we obtain
\begin{eqnarray}
\langle P^{ge}\rangle_{q} & = & \frac{4 V^2}{k_{B}T}\int\frac{d^{3}k^{\prime}}{(2\pi)^{3}}\int\frac{d^{3}k}{(2\pi)^{3}} P_{k,k^{\prime}}^{ge} \nonumber \\ 
  & = & \left(\frac{2\pi}{\hbar}\right)N^{2}\left(\epsilon_{F}\right)N_{imp}n_{g}(1-n_{e})\left[\frac{\beta}{e^{\beta\hbar\omega_{0}}-1}\right]g_{nr}^2,
\label{peg0} 
\end{eqnarray}
and
\begin{eqnarray}
\langle P^{eg}\rangle_{q} & = & \frac{4 V^2}{k_{B}T}\int\frac{d^{3}k^{\prime}}{(2\pi)^{3}}\int\frac{d^{3}k}{(2\pi)^{3}} P_{k,k^{\prime}}^{eg} \nonumber \\ 
  & = & \left(\frac{2\pi}{\hbar}\right)N^{2}\left(\epsilon_{F}\right)N_{imp}n_{e}(1-n_{g})\left[\frac{\beta}{1-e^{-\beta\hbar\omega_{0}}}\right]g_{nr}^2,
\label{peg}  
\end{eqnarray}
where
\be
g_{nr}^{2}=\frac{e^{2}}{\varepsilon_{0}^{2}}\frac{\mu^{2}}{3}\frac{\hbar\omega_{0}}{k_F^2}
\left[\frac{1}{1+2k_F^2\lambda_{TF}^2}+\ln{\left(1+2k_F^2\lambda_{TF}^2\right)}-1\right].
\label{gnr}
\ee
%
%
%
\begin{figure}
\begin{centering}
\includegraphics[width = 3.in]{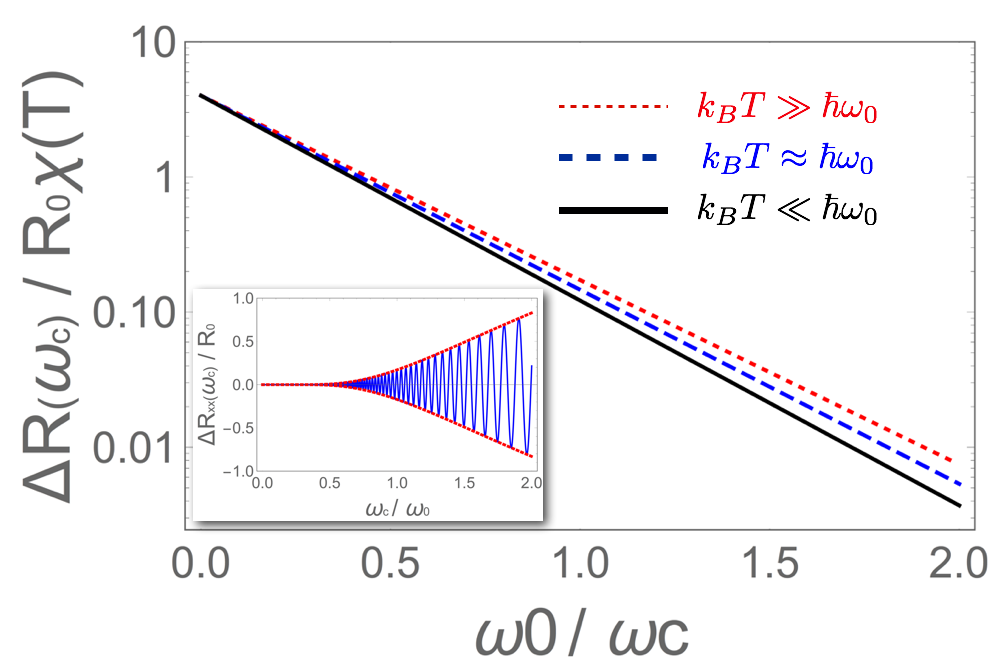}
\par\end{centering}
\caption{Main panel: amplitude of the Shubnikov-de-Haas oscillations for the case
of $\omega_0<\omega_c$ for different temperatures. Solid (black) line, for $k_B T\ll\hbar\omega_0$;
dashed (blue) line, for $k_B T\approx\hbar\omega_0$; and dotted (red) line,
for $k_B T\gg\hbar\omega_0$. Inset: Shubnikov-de-Haas oscillations at low magnetic fields, for the 
case of $\omega_c<\omega_0$.}
\label{Shubnikov-de-Haas}
\end{figure}
%

Remarkably, we now obtain that for the quantum lifetime, $1/\tau_{q}$, there exists a
one-to-one correspondence between the nonradiative emission and absorption rates, $\Gamma_{nr}^{e\rightarrow g}$
and $\Gamma_{nr}^{g\rightarrow e}$, and the inelastic contributions to electron-impurity scattering processes, 
$1/\tau_{eg}$ and $1/\tau_{ge}$. 
These results confirm our predictions that one can determine the nonradiative emission and absorption rates 
by means of magneto-transport observables such as, for example, the amplitude of the Shubnikov-de-Haas 
oscillations which is given by \cite{dingle}
\begin{equation}
\Delta R(\omega_c)=4 R_0 \chi(T) \; e^{-\pi/\omega_c\tau_q},
\end{equation}
which is governed by $\tau_{q}$ and it is shown in the main panel of Fig.~(\ref{Shubnikov-de-Haas}). 
From such a Dingle plot \cite{dingle}, we see that, 
already at small magnetic fields, $\omega_0/\omega_c\gg 1$, even the smallest contributions to the
quantum lifetime, $\tau_{q}$, from nonradiative transitions may lead to sizeble deviations 
of the amplitude of the Shubnikov-de-Haas oscillations from the pure elastic case at $k_B T\ll\hbar\omega_0$.
If we recall that, for $k_B T\leq\omega_0$ absorption processes dominate, as demonstrated
in Fig. (\ref{NR-Decay-Absorption-Rates}), we can promptly identify that the deviation of the dashed (blue) 
curve from the solid (black) curve in Fig. (\ref{Shubnikov-de-Haas}) is predominantly due to absorption 
processes, $\Gamma_{nr}^{g\rightarrow e}\gg \Gamma_{nr}^{e\rightarrow g}$. On the other hand, for the dotted (red) 
curve in Fig. (\ref{Shubnikov-de-Haas}), valid for $k_B T\gg\hbar\omega_0$, both emission and absorption
processes contribute equally, so that $\Gamma_{nr}^{g\rightarrow e}\approx\Gamma_{nr}^{e\rightarrow g}$. 
These findings suggest one can not only extract nonradiative decay rates from the Shubnikov-de-Haas oscillations via the quantum lifetime, but also detect the presence of the emitters inside a metallic system, as nonradiative decay rates strongly depend on temperature, as we previously demonstrate.  

\section{Conclusions}

In summary, we investigate nonradiative emission and absorption rates of two-level quantum
emitters embedded in a metal at low temperatures. Using Fermi's golden
rule, we derive expressions for both nonradiative transition rates, 
showing they are intrinsically related to electronic transport in the host metallic material. Indeed, we demonstrate nonradiative emission and absorption rates
could be directly determined by the knowledge of experimentally 
accessible transport quantities, such as the optical and ac-conductivity, and even the dc-resistivity. For concreteness, we consider the case of Shubnikov-de-Haas oscillations, governed by the quantum lifetime, which we demonstrate to be proportional to the nonradiative emission and absorption rates. Altogether our results not 
only provide a microscopic description of nonradiative decay channels in metals, but they also allows one to identify and 
differentiate them to other decay channels, which is crucial to understand and control light-matter interactions at the nanoscale.

\begin{acknowledgments}
We thank F.S.S. Rosa, C. Farina, and L.S. Menezes for fruitful discussions. We acknowledge CNPq, CAPES, and FAPERJ for financial support. F.A.P. also thanks the The Royal Society-Newton Advanced Fellowship (Grant no. NA150208) for financial support.
\end{acknowledgments}

\end{document}